\documentclass[uslettersize, 10pt, conference]{ieeeconf}      

\IEEEoverridecommandlockouts                              

\usepackage[OT1]{fontenc} 
\usepackage[english]{babel}
\usepackage{cite}
\usepackage[left=19.1mm,right=19.1mm,top=19.1mm,bottom=19.1mm]{geometry}
\usepackage{amsmath,amssymb,amsfonts}
\usepackage{textcomp}
\usepackage{xcolor}
\usepackage{blindtext}
\usepackage{subfiles} 
\usepackage{mathtools}
\usepackage{adjustbox}
\usepackage{tikz}
\usepackage{svg}
\usepackage{booktabs} 
\usepackage[final]{microtype} 
\usepackage{algorithm} 
\usepackage[algo2e,ruled,vlined]{algorithm2e} 
\usepackage{hyperref}
\hypersetup{hidelinks, breaklinks = true}
\usepackage[capitalise]{cleveref}
\usepackage{mathdots}
\usepackage{yhmath}
\usepackage{cancel}
\usepackage{color}
\usepackage{array}
\usepackage{multirow}
\usepackage{gensymb}
\usepackage{adjustbox}
\usepackage{tabularx}
\usepackage{comment}
\usepackage{extarrows}
\usepackage{booktabs}
\usepackage{diagbox}
\usepackage{graphicx}
\usepackage{subcaption}
\usepackage[normalem]{ulem}
\useunder{\uline}{\ul}{}
\usetikzlibrary{fadings}
\usetikzlibrary{patterns}
\usetikzlibrary{shadows.blur}
\usetikzlibrary{shapes}
\graphicspath{ {./Figures/} }
\usepackage{amsmath}
\usepackage{tikz}
\usepackage{mathdots}
\usepackage{yhmath}
\usepackage{cancel}
\usepackage{color}
\usepackage{array}
\usepackage{multirow}
\usepackage{amssymb}
\usepackage{bm}
\usepackage{gensymb}
\usepackage{tabularx}
\usepackage{extarrows}
\usepackage{booktabs}
\usetikzlibrary{fadings}
\usetikzlibrary{patterns}
\usetikzlibrary{shadows.blur}
\usetikzlibrary{shapes}
\usepackage{algcompatible}
\SetKwComment{Comment}{/* }{ */}

\newtheorem{defn}{Definition}[section]

\newtheorem{rem}{Remark}

\title{\LARGE \bf
Safe Optimal Interactions Between Automated and Human-Driven Vehicles in Mixed Traffic with Event-triggered Control Barrier Functions

\thanks{
This work was supported in part by NSF under grants CNS-2149511, ECCS-1931600,
DMS-1664644 and CNS-1645681, by ARPAE under grant DE-AR0001282,
and by the MathWorks.}
}

\author{
Anni Li, Christos G. Cassandras and Wei Xiao
\thanks{ A. Li and C. G. Cassandras are
with the Division of Systems Engineering and Center for Information and
Systems Engineering, Boston University, Brookline, MA 02446.
{\tt\small \{anlianni, cgc\}@bu.edu}}
\thanks{W. Xiao is with the Computer Science and Artificial Intelligence Lab, Massachusetts Institute of Technology. 
        {\tt\small weixy@mit.edu}}
}

\begin{document}

\maketitle
\thispagestyle{empty}
\pagestyle{empty}




\begin{abstract}
This paper studies safe driving interactions between Human-Driven Vehicles (HDVs) and Connected and Automated Vehicles (CAVs) in mixed traffic where the dynamics and control policies of HDVs are unknown and hard to predict.
In order to address this challenge, we employ event-triggered Control Barrier Functions (CBFs) to estimate the HDV model online, construct data-driven and state-feedback safety controllers, and transform constrained optimal control problems for CAVs into a sequence of event-triggered quadratic programs. We show that we can ensure collision-free between HDVs and CAVs and demonstrate the robustness and flexibility of our framework on different types of human drivers in lane-changing scenarios while guaranteeing safety with human-in-the-loop interactions. 
\end{abstract}

\section{INTRODUCTION}
Connected and Automated Vehicles (CAVs), also known as ``self-driving cars'', promise to significantly transform the operation of transportation networks and improve their performance by assisting drivers in making decisions so as to reduce accidents, as well as travel times, energy consumption, air pollution, and traffic congestion \cite{guanetti2018control,schrank2015urban,tideman2007review}. The cooperative control of CAVs has attracted a surge of interest in providing opportunities for vehicles to travel safely and optimally while enhancing the efficient operation of traffic networks \cite{rios2016survey,xu2019grouping}. 

However, 100\% CAV penetration is not likely in the near future, raising the question of how to benefit from the presence of at least some CAVs in mixed traffic and to still guarantee safety when CAVs must interact with uncontrollable Human-Driven Vehicles (HDVs) \cite{ghiasi2019mixed,wang2020controllability}. To address this challenge, efforts have concentrated on developing accurate car-following models, as in \cite{zhao2018optimal}, aiming at a deterministic quantification of HDV states, 
while \cite{mahbub2023safety} considers vehicle interactions, employs a prediction model to estimate HDV behaviors in real-time and directly controls CAVs to force HDVs to form platoons. In an effort to accurately model human driver behavior, a concept of social value orientation is defined in \cite{schwarting2019social} to characterize an agent's proclivity for social behavior or individualism and further predict human behavior. Considering vehicle interactions, a game-theoretic approach is used in \cite{wang2019game} to assist CAVs in evaluating the best possible response to an opponent's actions. 
While existing approaches have often shown impressive performance in managing vehicle interactions in mixed traffic, they assume known dynamics for HDVs.
However, the uncontrollable human behaviors under real-world conditions makes HDV models difficult to rely on for accurate predictions, which further increases the collision risk for CAVs and compromises their ability to guarantee safety. Moreover, most of the methods used are computationally expensive. The high complexity of obtaining accurate solutions motivates the use of Control Barrier Functions (CBFs) \cite{ames2014control,xiao2023safe,xiao2019control} to improve computation efficiency without compromising safety guarantees.

In recent years, numerous works use CBFs to enforce system safety, and employ Control Lyapunov Functions (CLFs) to make the system state converge to desired values \cite{ames2014control,ames2012control,wang2017safety,lopez2020robust}. The CBF method is usually sub-optimal, and one approach to address this limitation is to combine optimal control solutions with CBFs, leading to the Optimal control with CBFs (OCBF) approach \cite{xiao2021bridging}, in which reference trajectories are first obtained through Hamiltonian analysis in order to formulate tracking problems with quadratic objectives with the original safety constraints replaced and guaranteed by CBFs. Moreover, such optimal tracking problems can be solved by discretizing time and transforming them into a sequence of Quadratic Programs (QPs) at each time step with the assumption that control is a constant during each such time interval. This assumption gives rise to the problem that each time discretization interval needs to be sufficiently small to ensure the feasibility of each QP at any one time step. One way to solve this problem is to adopt \emph{event-triggered} approaches as proposed in  \cite{tabuada2007event} and further in \cite{xiao2022event} to deal with unknown system dynamics. However, there has been little consideration of human factors in conjunction with CBFs.

In this paper, we study safe driving interactions between CAVs and HDVs in a mixed-traffic environment, in which case the dynamics and human control policies of HDVs are unknown. We
 adopt the event-triggered CBF method proposed in \cite{xiao2022event} for CAVs to ensure the safety between CAVs and HDVs, and implement it in highway lane-changing maneuvers as shown in Fig. \ref{fig:lane_change_process}: in this case, the green vehicles 1 and $C$ are assumed to be cooperating CAVs, the red vehicle $H$ is an uncontrollable HDV, and the gray vehicle $U$ is considered as a dynamic obstacle moving at a slower speed than CAVs. This motivates $C$
to try and overtake $U$ so as to jointly minimize its travel time and energy consumption while ensuring a small speed deviation from the fast lane traffic flow in order to minimize any disruption caused by the maneuver. As shown in \cite{armijos2022cooperative}, a key step for $C$ to perform an optimal lane-changing maneuver is to optimally choose a  pair on the fast lane to merge in between; this can be achieved by cooperating CAVs in a 100\% CAV penetration environment. In mixed traffic, due to the existence of HDVs, minimizing the time, energy, and speed deviations in the fast lane flow is no longer ensured. This problem is addressed in \cite{li2023cooperative}, which considers vehicle interactions so as to design the best response for a CAV to actions by its neighboring HDVs, and further provides an option for CAV $C$ to merge ahead of CAV 1 so as to eliminate any dependence on the HDV involved in the maneuver. However, an HDV's behavior still needs to be estimated by any given model, and the robustness of the control cannot be guaranteed in practice, given the uncertainties included in HDV models when $C$ decides to merge ahead of $H$. Moreover, the optimal planning of the maneuver is limited in the longitudinal direction.

The above issues are resolved in this work which provides enhanced robustness of the optimal policies to uncontrollable HDVs in mixed traffic. The main contributions of this paper are summarized as follows
\begin{itemize}

    \item We propose a safe and robust
    human interaction framework in mixed traffic using event-triggered CBFs under the case of unknown (generally nonlinear, but affine in the control) HDV dynamics and unpredictable human-in-the-loop control policies.
    
    \item Both longitudinal and lateral maneuvers are combined together in a lane-changing maneuver with an ellipsoidal safety region determined for vehicles so as to guarantee safety in a 2D manner during the entire maneuver.  

    \item The optimal pair for the lane-changing CAV to merge in between is determined in real time, depending on the aggressiveness of HDVs.

    \item We demonstrate safe human interactions on different types of human drivers (e.g., aggressive, hesitant, conservative) in mixed traffic highway lane merging.
    
\end{itemize}

\section{PRELIMINARIES}
\begin{defn}
(\emph{Class $\mathcal{K}$ Function} \cite{khalil2002nonlinear}) A continuous function $\alpha:[0,a)\rightarrow [0,\infty),a>0$ is said to belong to class $\mathcal{K}$ if it is strictly increasing and $\alpha(0)=0$.
\end{defn}

Consider an affine control system
\begin{equation}
\label{eq:affine_system}
    \dot{\bm{x}} = f(\bm{x}) + g(\bm{x})
    \bm{u},
\end{equation}
where $\bm{x}\in \mathbb{R}^n, \bm{u}\in \mathcal{U}\subset \mathbb{R}^q$ denote the state and control vector respectively, $f:\mathbb{R}^n \rightarrow \mathbb{R}^n$ and $g:\mathbb{R}^n \rightarrow \mathbb{R}^{n\times q}$ are Lipschitz continuous. 

\begin{defn}
\label{def:Forward_Invariant_Set}
    (\emph{Forward Invariant Set}) A set $C\subset \mathbb{R}^n$ is forward invariant for system \ref{eq:affine_system} if its solutions starting at any $\bm{x}(0)\in C$ satisfy $\bm{x}(t)\in C$ for $\forall t\geq 0$.
\end{defn}

\begin{defn}
    (\emph{Control Barrier Function }\cite{ames2014control}) Given a set $C$ as in Def. \ref{def:Forward_Invariant_Set}, $b(\bm{x})$ is a Control Barrier Function (CBF) for system (\ref{eq:affine_system}) if there exists a class $\mathcal{K}$ function $\alpha$ such that 
    \begin{equation}
    \label{eq:CBF_def}
        \sup_{\bm{u}\in \mathcal{U}} [L_fb(\bm{x})+L_gb(\bm{x})\bm{u}+\alpha(b(\bm{x}))]\geq 0, ~ \forall x\in C,
    \end{equation}
    where $L_f,L_g$ denote the Lie derivatives along $f$ and $g$, respectively. It is assumed that $L_gb(\bm{x})\ne 0$ when $b(\bm{x})=0$.
\end{defn}

\begin{defn}
    (\emph{Control Lyapunov Function} \cite{ames2012control}) A continuously differentiable function $V:\mathbb{R}^n \rightarrow \mathbb{R}$ is an exponentially stabilizing Control Lyapunov Function (CLF) for system (\ref{eq:affine_system}) if there exists constants $c_1>0,c_2>0,c_3>0$ such that for all $\bm{x}\in \mathbb{R}^n$, $c_1 ||\bm{x}||^2\leq V(\bm{x})\leq c_2||\bm{x}||^2$,
    \vspace{-1mm}
    \begin{equation} \label{eqn:clf}
        \inf_{\bm{u}\in \mathcal{U}} [L_fV(\bm{x})+L_gV(\bm{x})\bm{u}+c_3V(\bm{x})] \leq 0.
    \end{equation}
\end{defn}

Many existing works \cite{ames2014control,xiao2023safe,xiao2019decentralized,sabouni2023optimal} replace the safety constraints with CBFs in the optimization problem to map state constraints to state-dependent control constraints so as to reduce the computational complexity, and still guarantee their enforcement at the expense of some loss in performance.

\section{PROBLEM FORMULATION}
\begin{figure} [hpt]
    \centering
    \vspace*{-\baselineskip}
    \begin{adjustbox}{width=0.8\linewidth, height = 2.5cm,center}

\tikzset{every picture/.style={line width=0.75pt}} 

\begin{tikzpicture}[x=0.75pt,y=0.75pt,yscale=-1,xscale=1]

\draw [line width=3]    (16,55.54) -- (354.76,55.7) -- (654,55.54) ;
\draw [line width=3]    (18,260.54) -- (317.62,259.03) -- (650,257.54) ;
\draw [color={rgb, 255:red, 155; green, 155; blue, 155 }  ,draw opacity=1 ][fill={rgb, 255:red, 248; green, 231; blue, 28 }  ,fill opacity=1 ][line width=3]  [dash pattern={on 7.88pt off 4.5pt}]  (19,158.54) -- (646,157.54) ;
\draw [color={rgb, 255:red, 245; green, 166; blue, 35 }  ,draw opacity=1 ][fill={rgb, 255:red, 248; green, 231; blue, 28 }  ,fill opacity=1 ][line width=0.75]  [dash pattern={on 4.5pt off 4.5pt}]  (25,209.54) -- (639,210.54) ;
\draw [color={rgb, 255:red, 245; green, 166; blue, 35 }  ,draw opacity=1 ][fill={rgb, 255:red, 248; green, 231; blue, 28 }  ,fill opacity=1 ][line width=0.75]  [dash pattern={on 4.5pt off 4.5pt}]  (21,111.54) -- (644,112.54) ;
\draw (148.62,113.62) node  {\includegraphics[width=135.57pt,height=119.13pt]{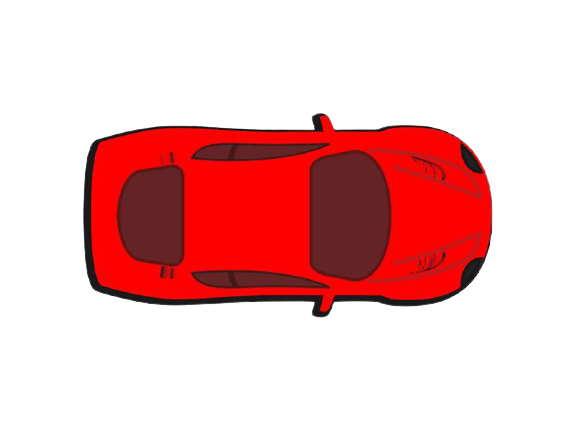}};
\draw (444.09,111.63) node  {\includegraphics[width=91.22pt,height=57.34pt]{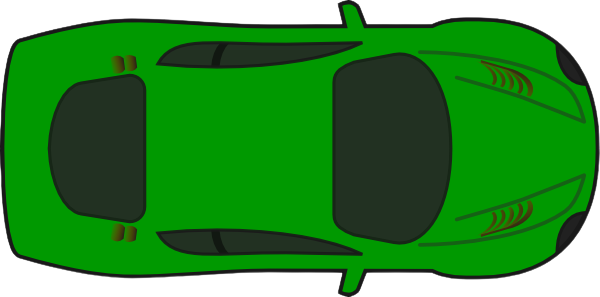}};
\draw   (30.74,85.54) -- (35.37,60.79) -- (40,85.54) -- (37.68,85.54) -- (37.68,135.04) -- (40,135.04) -- (35.37,159.79) -- (30.74,135.04) -- (33.05,135.04) -- (33.05,85.54) -- cycle ;
\draw  [fill={rgb, 255:red, 0; green, 0; blue, 0 }  ,fill opacity=1 ] (228.44,173.22) .. controls (228.2,172.37) and (228.7,171.49) .. (229.55,171.25) -- (252.8,164.8) .. controls (253.65,164.57) and (254.53,165.06) .. (254.76,165.91) -- (256.35,171.63) .. controls (256.59,172.48) and (256.09,173.36) .. (255.24,173.6) -- (231.99,180.05) .. controls (231.14,180.29) and (230.26,179.79) .. (230.03,178.94) -- cycle ;
\draw  [fill={rgb, 255:red, 0; green, 0; blue, 0 }  ,fill opacity=1 ] (243.44,233.22) .. controls (243.2,232.37) and (243.7,231.49) .. (244.55,231.25) -- (267.8,224.8) .. controls (268.65,224.57) and (269.53,225.06) .. (269.76,225.91) -- (271.35,231.63) .. controls (271.59,232.48) and (271.09,233.36) .. (270.24,233.6) -- (246.99,240.05) .. controls (246.14,240.29) and (245.26,239.79) .. (245.03,238.94) -- cycle ;
\draw  [fill={rgb, 255:red, 0; green, 0; blue, 0 }  ,fill opacity=1 ] (304.77,227.42) .. controls (304.23,226.72) and (304.36,225.71) .. (305.06,225.18) -- (324.16,210.45) .. controls (324.86,209.91) and (325.86,210.04) .. (326.4,210.74) -- (330.02,215.44) .. controls (330.56,216.14) and (330.43,217.14) .. (329.73,217.68) -- (310.63,232.41) .. controls (309.93,232.94) and (308.93,232.81) .. (308.39,232.12) -- cycle ;
\draw  [fill={rgb, 255:red, 0; green, 0; blue, 0 }  ,fill opacity=1 ] (290.77,162.42) .. controls (290.23,161.72) and (290.36,160.71) .. (291.06,160.18) -- (310.16,145.45) .. controls (310.86,144.91) and (311.86,145.04) .. (312.4,145.74) -- (316.02,150.44) .. controls (316.56,151.14) and (316.43,152.14) .. (315.73,152.68) -- (296.63,167.41) .. controls (295.93,167.94) and (294.93,167.81) .. (294.39,167.12) -- cycle ;
\draw (290.09,193.17) node [rotate=-348.35] {\includegraphics[width=91.22pt,height=57.34pt]{plots/greenVehicleTopView.png}};
\draw [color={rgb, 255:red, 208; green, 2; blue, 27 }  ,draw opacity=1 ][line width=2.25]  [dash pattern={on 6.75pt off 4.5pt}]  (180.2,215.87) -- (374.2,215.87) ;
\draw [color={rgb, 255:red, 208; green, 2; blue, 27 }  ,draw opacity=1 ][line width=2.25]  [dash pattern={on 6.75pt off 4.5pt}]  (180.2,215.87) -- (381.28,174.67) ;
\draw [shift={(385.2,173.87)}, rotate = 168.42] [color={rgb, 255:red, 208; green, 2; blue, 27 }  ,draw opacity=1 ][line width=2.25]    (17.49,-5.26) .. controls (11.12,-2.23) and (5.29,-0.48) .. (0,0) .. controls (5.29,0.48) and (11.12,2.23) .. (17.49,5.26)   ;
\draw [color={rgb, 255:red, 65; green, 117; blue, 5 }  ,draw opacity=1 ][line width=1.5]    (346.2,179.87) -- (374.09,146.02) ;
\draw [shift={(376,143.7)}, rotate = 129.49] [color={rgb, 255:red, 65; green, 117; blue, 5 }  ,draw opacity=1 ][line width=1.5]    (14.21,-4.28) .. controls (9.04,-1.82) and (4.3,-0.39) .. (0,0) .. controls (4.3,0.39) and (9.04,1.82) .. (14.21,4.28)   ;
\draw  [draw opacity=0] (353.69,170.05) .. controls (355.37,171.1) and (356.48,172.91) .. (356.49,174.96) .. controls (356.5,176.75) and (355.67,178.36) .. (354.35,179.44) -- (350.25,174.99) -- cycle ; \draw   (353.69,170.05) .. controls (355.37,171.1) and (356.48,172.91) .. (356.49,174.96) .. controls (356.5,176.75) and (355.67,178.36) .. (354.35,179.44) ;  
\draw  [draw opacity=0] (218,209.02) .. controls (219.43,210.1) and (220.33,211.72) .. (220.34,213.53) .. controls (220.34,214.28) and (220.19,215) .. (219.91,215.66) -- (213.79,213.56) -- cycle ; \draw   (218,209.02) .. controls (219.43,210.1) and (220.33,211.72) .. (220.34,213.53) .. controls (220.34,214.28) and (220.19,215) .. (219.91,215.66) ;  
\draw  [dash pattern={on 0.84pt off 2.51pt}]  (233,151.7) -- (297,135.7) ;
\draw [shift={(297,135.7)}, rotate = 165.96] [color={rgb, 255:red, 0; green, 0; blue, 0 }  ][line width=0.75]    (0,5.59) -- (0,-5.59)   ;
\draw [shift={(233,151.7)}, rotate = 165.96] [color={rgb, 255:red, 0; green, 0; blue, 0 }  ][line width=0.75]    (0,5.59) -- (0,-5.59)   ;
\draw  [color={rgb, 255:red, 208; green, 2; blue, 27 }  ,draw opacity=1 ][dash pattern={on 4.5pt off 4.5pt}] (46,109.08) .. controls (46,81.17) and (91.89,58.55) .. (148.5,58.55) .. controls (205.11,58.55) and (251,81.17) .. (251,109.08) .. controls (251,136.99) and (205.11,159.62) .. (148.5,159.62) .. controls (91.89,159.62) and (46,136.99) .. (46,109.08) -- cycle ;
\draw  [color={rgb, 255:red, 208; green, 2; blue, 27 }  ,draw opacity=1 ][dash pattern={on 4.5pt off 4.5pt}] (347,109.08) .. controls (347,81.17) and (392.89,58.55) .. (449.5,58.55) .. controls (506.11,58.55) and (552,81.17) .. (552,109.08) .. controls (552,136.99) and (506.11,159.62) .. (449.5,159.62) .. controls (392.89,159.62) and (347,136.99) .. (347,109.08) -- cycle ;
\draw  [color={rgb, 255:red, 208; green, 2; blue, 27 }  ,draw opacity=1 ][dash pattern={on 4.5pt off 4.5pt}] (183.8,221.79) .. controls (176.47,194.87) and (214.8,160.98) .. (269.42,146.11) .. controls (324.05,131.24) and (374.27,141.01) .. (381.6,167.94) .. controls (388.93,194.87) and (350.6,228.76) .. (295.98,243.63) .. controls (241.35,258.5) and (191.13,248.72) .. (183.8,221.79) -- cycle ;
\draw (511.69,204.53) node [rotate=-0.02,xslant=0] {\includegraphics[width=105.52pt,height=52.48pt]{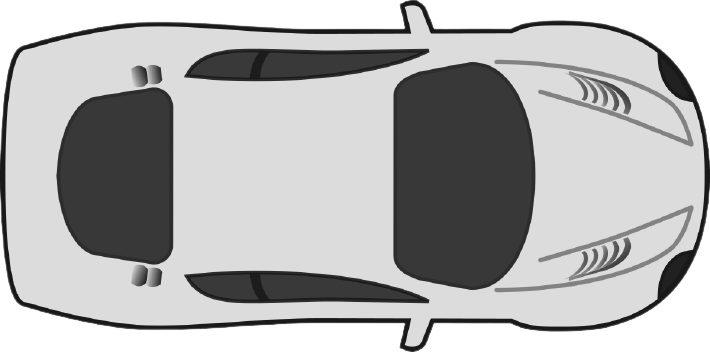}};

\draw (453.42,113.36) node  [font=\large,color={rgb, 255:red, 255; green, 255; blue, 255 }  ,opacity=1 ] [align=left] {\begin{minipage}[lt]{42.95pt}\setlength\topsep{0pt}
1
\end{minipage}};
\draw (162.45,118.46) node  [font=\large,color={rgb, 255:red, 255; green, 255; blue, 255 }  ,opacity=1 ] [align=left] {\begin{minipage}[lt]{47.72pt}\setlength\topsep{0pt}
H 
\end{minipage}};
\draw (18,85.4) node [anchor=north west][inner sep=0.75pt]  [font=\large]  {$l$};
\draw (281.77,179.72) node  [font=\large,color={rgb, 255:red, 255; green, 255; blue, 255 }  ,opacity=1 ] [align=left] {\begin{minipage}[lt]{17.44pt}\setlength\topsep{0pt}
C
\end{minipage}};
\draw (358.2,155.74) node [anchor=north west][inner sep=0.75pt]  [font=\large]  {$\phi $};
\draw (207,187.24) node [anchor=north west][inner sep=0.75pt]  [font=\large]  {$\theta $};
\draw (369.2,181.24) node [anchor=north west][inner sep=0.75pt]  [font=\large,color={rgb, 255:red, 208; green, 2; blue, 27 }  ,opacity=1 ]  {$v( t)$};
\draw (247,120.39) node [anchor=north west][inner sep=0.75pt]    {$L_{w}$};
\draw (517.42,208.36) node  [font=\large,color={rgb, 255:red, 0; green, 0; blue, 0 }  ,opacity=1 ] [align=left] {\begin{minipage}[lt]{42.95pt}\setlength\topsep{0pt}
U
\end{minipage}};
\draw (13,193.4) node [anchor=north west][inner sep=0.75pt]  [font=\Large]  {$y=0$};

\end{tikzpicture}

    \end{adjustbox}
    \caption{\small The basic lane-changing maneuver process. The red vehicle is an HDV, green vehicles are CAVs, and the grey vehicle is a slow-moving and uncontrollable vehicle.}
    \label{fig:lane_change_process}
\end{figure}

The lane change maneuver shown in Fig. \ref{fig:lane_change_process} is triggered by CAV $C$ when an obstacle (e.g., slow-moving vehicle $U$) ahead is detected. In general, such a maneuver can be initiated at any arbitrary time set by the CAV. 
The framework proposed in this paper can be used in any conflict area involving vehicle interactions, but we limit ourselves to this lane-changing setting which we view as the most challenging among them.
We aim to minimize the maneuver time and energy expended, while alleviating any disruption to the fast lane traffic flow. Moreover, considering the presence of HDVs, $C$ also needs to be aware of the behavior of its surrounding HDVs in order to guarantee safety.

\textbf{Vehicle Dynamics.} The dynamics and control policy of the HDV are unknown in this case. Assume the slow vehicle $U$ keeps traveling in the slow lane with a constant speed $v_U$. For each CAV in Fig. \ref{fig:lane_change_process}, indexed by $i\in \{1,C\}$, its dynamics take the form
\begin{equation}
\label{eq:CAVs_dynamics}
\begin{adjustbox}{width=0.9\linewidth}
    $\underbrace{\left[\begin{array}{c}
    \dot{x}_i \\
    \dot{y}_i \\
    \dot{\theta}_i \\
     \dot{v}_i   \end{array}\right]}_{\bm{\dot{x}}_i}$
    =
    $\underbrace{\left[\begin{array}{c}
    v_i \cos \theta_i \\
    v_i \sin \theta_i \\
    0 \\
    0    \end{array}\right]}_{f\left(\bm{x}_i(t)\right)}$+
    $\underbrace{\left[\begin{array}{cc}
    0 & -v_i \sin \theta_i \\
    0 & v_i \cos \theta_i \\
    0 & v_i / L_w \\
    1 & 0    \end{array}\right]}_{g\left(\bm{x}_i(t)\right)}
    \underbrace{\left[\begin{array}{l}
    u_i \\
    \phi_i
    \end{array}\right]}_{\bm{u}_i(t)}$
    \end{adjustbox}
\end{equation}
where $x_i(t),y_i(t),\theta_i(t),v_i(t)$ represent the current longitudinal position, lateral position, heading angle, and speed, respectively. $u_i(t)$ and $\phi_i(t)$ are the acceleration and steering angle (controls) of vehicle $i$ at time $t$, respectively, $g(\bm{x}_i(t))=[g_u(\bm{x}_i(t)),g_{\phi}(\bm{x}_i(t))]$. The maneuver starts at time $t_0$ and ends at time $t_f$ when $C$ has completely switched to the target lane. The control input and speed for all vehicles are constrained as follows:

$\vspace{-2mm}$
{\small
\begin{align}
\label{eq:uv_constraints}
    \bm u_{i_{\min}}\leq \bm u_i(t)\leq \bm u_{i_{\max}},~v_{i_{\min}}\leq v_i(t)\leq v_{i_{\max}}, \; i\in\{1,C\},
\end{align}}

\noindent where $\bm u_{i_{\min}},\bm u_{i_{\max}}\in\mathbb{R}^2$ denote the minimum and maximum control bounds for vehicle $i$, respectively. $v_{i_{\min}}>0$ and $v_{i_{\max}}>0$ are vehicle $i$'s allowable minimum and maximum speed. 
Setting $l$ as the width of the road, $y=0$ axis is the center of the slow lane in Fig. \ref{fig:lane_change_process}, we have $y_C(t_0)=0$,
and the lateral positions of vehicles satisfy
\begin{align}
\label{eq:ylimits}
    -\frac{l}{2} \leq y_i(t) \leq \frac{3}{2}l,~ i\in\{1,C\}.
\end{align}

\begin{figure} [hpt]
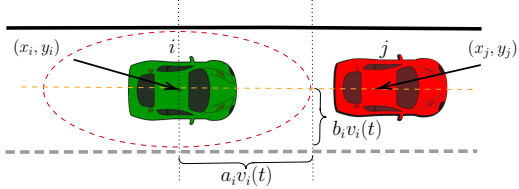

    \centering
    \vspace*{-\baselineskip}
    \begin{adjustbox}{width=0.8\linewidth, height = 2.5cm,center}

\tikzset{every picture/.style={line width=0.75pt}} 

\begin{tikzpicture}[x=0.75pt,y=0.75pt,yscale=-1,xscale=1]

\draw  [color={rgb, 255:red, 208; green, 2; blue, 27 }  ,draw opacity=1 ][dash pattern={on 4.5pt off 4.5pt}] (96,135.95) .. controls (96,100.23) and (163.6,71.28) .. (247,71.28) .. controls (330.4,71.28) and (398,100.23) .. (398,135.95) .. controls (398,171.66) and (330.4,200.62) .. (247,200.62) .. controls (163.6,200.62) and (96,171.66) .. (96,135.95) -- cycle ;
\draw (253.09,137.63) node  {\includegraphics[width=91.22pt,height=57.34pt]{plots/greenVehicleTopView.png}};
\draw (487.62,134.62) node  {\includegraphics[width=135.57pt,height=119.13pt]{plots/redVehicleTopView.png}};
\draw   (249,212.28) .. controls (249.06,216.95) and (251.42,219.25) .. (256.09,219.19) -- (314.59,218.41) .. controls (321.26,218.32) and (324.62,220.61) .. (324.69,225.28) .. controls (324.62,220.61) and (327.92,218.24) .. (334.59,218.15)(331.59,218.19) -- (393.09,217.37) .. controls (397.76,217.31) and (400.06,214.95) .. (400,210.28) ;
\draw  [dash pattern={on 0.84pt off 2.51pt}]  (399.5,35.95) -- (399.72,81.29) -- (400.5,237.95) ;
\draw [line width=3]    (54,67.28) -- (353.76,66.7) -- (591,68.28) ;
\draw [color={rgb, 255:red, 155; green, 155; blue, 155 }  ,draw opacity=1 ][fill={rgb, 255:red, 248; green, 231; blue, 28 }  ,fill opacity=1 ][line width=3]  [dash pattern={on 7.88pt off 4.5pt}]  (53,206.28) -- (586,205.28) ;
\draw [color={rgb, 255:red, 245; green, 166; blue, 35 }  ,draw opacity=1 ][fill={rgb, 255:red, 248; green, 231; blue, 28 }  ,fill opacity=1 ][line width=0.75]  [dash pattern={on 4.5pt off 4.5pt}]  (71,133.28) -- (587,136.28) ;
\draw  [dash pattern={on 0.84pt off 2.51pt}]  (248.5,35.95) -- (249.5,237.95) ;
\draw   (401,198.28) .. controls (405.67,198.35) and (408.04,196.06) .. (408.11,191.39) -- (408.34,176.89) .. controls (408.45,170.22) and (410.83,166.93) .. (415.5,167) .. controls (410.83,166.93) and (408.55,163.56) .. (408.66,156.89)(408.61,159.89) -- (408.89,142.39) .. controls (408.96,137.72) and (406.67,135.35) .. (402,135.28) ;
\draw [color={rgb, 255:red, 0; green, 0; blue, 0 }  ,draw opacity=1 ][line width=1.5]    (130,102.77) -- (244.11,135.13) ;
\draw [shift={(247,135.95)}, rotate = 195.83] [color={rgb, 255:red, 0; green, 0; blue, 0 }  ,draw opacity=1 ][line width=1.5]    (14.21,-4.28) .. controls (9.04,-1.82) and (4.3,-0.39) .. (0,0) .. controls (4.3,0.39) and (9.04,1.82) .. (14.21,4.28)   ;
\draw [line width=1.5]    (590,108.77) -- (477.93,134.11) ;
\draw [shift={(475,134.77)}, rotate = 347.26] [color={rgb, 255:red, 0; green, 0; blue, 0 }  ][line width=1.5]    (14.21,-4.28) .. controls (9.04,-1.82) and (4.3,-0.39) .. (0,0) .. controls (4.3,0.39) and (9.04,1.82) .. (14.21,4.28)   ;

\draw (295,220.4) node [anchor=north west][inner sep=0.75pt]  [font=\LARGE]  {$a_{i} v_{i}( t)$};
\draw (421,168.4) node [anchor=north west][inner sep=0.75pt]  [font=\LARGE]  {$b_{i} v_{i}( t)$};
\draw (60,78.4) node [anchor=north west][inner sep=0.75pt]  [font=\Large]  {$( x_{i} ,y_{i})$};
\draw (236,78.4) node [anchor=north west][inner sep=0.75pt]  [font=\LARGE,color={rgb, 255:red, 0; green, 0; blue, 0 }  ,opacity=1 ]  {$i$};
\draw (474.38,79.89) node [anchor=north west][inner sep=0.75pt]  [font=\LARGE,color={rgb, 255:red, 0; green, 0; blue, 0 }  ,opacity=1 ]  {$j$};
\draw (574,79.4) node [anchor=north west][inner sep=0.75pt]  [font=\Large]  {$( x_{j} ,y_{j})$};

\end{tikzpicture}

    \end{adjustbox}
    \caption{Definition of an elliptical safe region.}
    \label{fig:safe_region}
\end{figure}

\textbf{Safety Constraints.} Similar to the longitudinal safe distance described in \cite{xiao2023safe}, we define an ellipsoidal safe region $b_{i,j}(\bm{x}_i,\bm{x}_j)$ for vehicles $i$ and $j$ during the entire maneuver:
\begin{equation}
\label{eq:safety_distance}
    b_{i,j}:= \dfrac{(x_j(t)-x_i(t))^2}{(a_i v_i(t))^2}+ \dfrac{(y_j(t)-y_i(t))^2}{(b_i v_i(t))^2} -1 \geq 0,
\end{equation}
where $j$ is $i$'s neighboring vehicle, $a_i,b_i$ are weights adjusting the length of the major and minor axes of the ellipse shown in Fig. \ref{fig:safe_region}, and the size of the safe region depends on speed. Note that $b_{i,j}$ is specified from the center of vehicle $i$ to the center of $j$. In other words, the center of vehicle $j$ must remain outside of $i$'s safe region during the entire maneuver. Defining an elliptical safe region considers the 2D safe distance between two vehicles. Since \eqref{eq:safety_distance} depends on speed, its CBF constraint only has relative degree one (i.e., we only need to take the derivative of the safety constraint along the dynamics once until the control explicitly shows in the derivative), implying lower complexity in CBF design.  

Therefore, CAV $i\in\{1,C\}$ in Fig. \ref{fig:lane_change_process} must satisfy the following constraints to guarantee safety during any lane change maneuver:

\begin{subequations}
\small
\label{eq:safety_constraints_original}
    \begin{align}
    \label{eq:bCH}
        & b_{C,H}\!=\! \dfrac{(x_C(t)\!-\!x_H(t))^2}{a_C^2}\!+\! \dfrac{(y_C(t)\!-\!y_H(t))^2}{b_C^2} - v_C^2(t)\geq0,\\
        \label{eq:b1C}
        & b_{1,C}= \dfrac{(x_1(t)\!-\!x_C(t))^2}{a_1^2}\!+\! \dfrac{(y_1(t)\!-\!y_C(t))^2}{b_1^2} -v_1^2(t)\geq 0, \\
        \label{eq:b1H}
        & b_{1,H}= \dfrac{(x_1(t)\!-\!x_H(t))^2}{a_1^2}\!+\! \dfrac{(y_1(t)\!-\!y_H(t))^2}{b_1^2} - v_1^2(t)\geq 0,\\
        \label{eq:bUC}
        & b_{U,C}= \dfrac{(x_U(t)\!-\!x_C(t))^2}{a_C^2}\!+\! \dfrac{(y_U(t)\!-\!y_C(t))^2}{b_C^2} - v_C^2(t)\geq 0,
    \end{align}
\end{subequations}
Each constraint in \eqref{eq:safety_constraints_original} ensures that the safe region of CAVs 1 or $C$ is not invaded by surrounding vehicles depicted in Fig. \ref{fig:lane_change_process}. For instance, \eqref{eq:bCH} necessitates that CAV $C$ maintains a safe distance from the HDV, such that the HDV remains exterior to the defined elliptical safe region.

\textbf{Optimal Control Problem Formulation.}
Our goal is to determine the optimal control policy for CAV $C$ to perform a safe lane change maneuver. The objective is to jointly minimize $C$'s energy consumption and speed deviation from traffic flow while guaranteeing safety. Considering cooperations between $C$ and 1, the joint cooperative optimal control problem (OCP) for both CAVs is given by:
\begin{align}
\nonumber 
        &\min\limits_{u_C(t),u_1(t),t_f} \int_{t_0}^{t_f} \dfrac{\alpha_u}{2}(u_C^2(t)+u_1^2(t))dt + \alpha_l(y_C(t_f)-l)^2\\  
        \label{eq:ocp}&\;\;\;\;\;\;\;\;\;\;\;\;\;\;\;\;\;\;\;\;+ \alpha_v[(v_C(t_f)-v_d)^2 + (v_1(t_f)-v_d)^2]\\
        \nonumber
        &\;\;\;\;\;\;\;\;\;\;\;\;\; s.t. ~ ~ (\ref{eq:CAVs_dynamics}),~(\ref{eq:uv_constraints}),~(\ref{eq:ylimits}),~(\ref{eq:safety_constraints_original})
\end{align}
where $v_d$ denotes the desired speed of CAVs in the fast lane, $\alpha_u,\alpha_l,\alpha_v$ are adjustable non-negative (properly normalized) weights for energy, desired lateral position, and desired speed, respectively. The CAV dynamics are given in (\ref{eq:CAVs_dynamics}) with state and control limits as in (\ref{eq:uv_constraints}) and (\ref{eq:ylimits}). Safety distances between all vehicle pairs in Fig. \ref{fig:lane_change_process} are constrained through (\ref{eq:safety_constraints_original}), requiring state knowledge of all vehicles. However, since HDVs are uncontrollable and unknown to CAVs in actuality, coupling the unknown HDV states with CAVs in safety constraints (\ref{eq:bCH}) and (\ref{eq:b1H}) makes (\ref{eq:ocp}) directly unsolvable.

Therefore, in this paper, we employ event-triggered CBFs \cite{xiao2022event} to solve (\ref{eq:ocp}). The critical step to solve OCP (\ref{eq:ocp}) using event-triggered CBFs is to first estimate HDV dynamics, which requires time discretization in order to update states at each time step so as to reduce the estimation error and obtain a more accurate model. Then, at each time step, CAVs can proceed with their maneuvers based on HDV state estimates, and replace the safety constraints in \eqref{eq:safety_constraints_original} with the event-triggered CBF constraints to enforce their satisfaction while reducing the computational complexity \cite{xiao2023safe} 
(the details of this replacement will be discussed in the following section).
Then, the problem \eqref{eq:ocp} can be transformed into a sequence of Quadratic Programs (QPs) similar to \cite{ames2014control}. 
Finally, we implement an \emph{event-driven} approach in the CBF-based QPs to find the next triggering time to solve the QP rather than relying on the time-driven method which requires a proper selection of discretization time steps sizes. The details of the overall framework are described in the following section.

\section{HUMAN-IN-THE-LOOP SAFE LANE-CHANGING MANEUVER}
\subsection{CBFs with Adaptive Dynamics for HDVs}
For CAV $C$ to safely and optimally change lanes, it has to rely on accurate HDV dynamics, which are difficult to model and predict in actual system operation. In this section, we introduce adaptively updated dynamics for the HDV based on real-time measurements so as to approximate its actual dynamics. We then show how to define CBFs based on these adaptive dynamics.

We begin by assuming adaptive nonlinear (but affine in the control) dynamics for the HDV:
\begin{equation}
\label{eq:hdv_nominal}
    \dot{\Bar{\bm{x}}}_H = f_a(\Bar{\bm{x}}_H) + g_a(\Bar{\bm{x}}_H)\bm{u}_H,
\end{equation}
where $f_a:\mathbb{R}^n \rightarrow \mathbb{R}$, $g_a:\mathbb{R}^n \rightarrow \mathbb{R}^{n\times q}$ are adaptive functions to accommodate the real unknown HDV dynamics, $u_H\in\mathbb{R}^q$ is the control input of HDV, and $\Bar{\bm{x}}_H\in X \subset \mathbb{R}^n$ is the estimated state vector corresponding to the real HDV states $\bm{x}_H$. Let
\begin{align}
\label{eq:error}
    \bm{e}:= \bm{x}_H - \Bar{\bm{x}}_H,
\end{align}
be the direct measurement error between the real HDV states $\bm{x}_H$ and estimated states $\Bar{\bm{x}}_H$. Then for any safety function $b_{i,H}(\bm{x}_i,\bm{x}_H)$ defined in \eqref{eq:safety_constraints_original} 
between vehicles $i$ and $H$, it has to satisfy 
\begin{align}
    b_{i,H}(\bm{x}_i,\bm{x}_H) = b_{i,H}(\bm{x}_i,\Bar{\bm{x}}_H+\bm{e}) \geq 0.
\end{align}
Differentiating $b_{i,H}(\bm{x}_i,\bm{x}_H)$, we have
{\small
\begin{align}
\nonumber
    \dfrac{db_{i,H}(\bm{x}_i,\bm{x}_H)}{dt} &= \dfrac{\partial b_{i,H}(\bm{x}_i,\Bar{\bm{x}}_H+\bm{e})}{\partial\bm{x}_i}\dot{\bm{x}}_i + \\
    \nonumber
    &\dfrac{\partial b_{i,H}(\bm{x}_i,\Bar{\bm{x}}_H+\bm{e})}{\partial\Bar{\bm{x}}_H}\dot{\Bar{\bm{x}}}_H + 
    \dfrac{\partial b_{i,H}(\bm{x}_i,\Bar{\bm{x}}_H+\bm{e})}{\partial\bm{e}}\dot{\bm{e}}
\end{align}}

Equivalent to the CBF constraint defined in \eqref{eq:CBF_def}, we have the CBF constraint to enforce safety for the unknown-dynamics HDV in the form

{\small
\begin{align}
\nonumber
    &\dfrac{\partial b_{i,H}(\bm{x}_i,\Bar{\bm{x}}_H+\bm{e})}{\partial\bm{x}_i}\dot{\bm{x}}_i + 
    \dfrac{\partial b_{i,H}(\bm{x}_i,\Bar{\bm{x}}_H+\bm{e})}{\partial\Bar{\bm{x}}_H}\dot{\Bar{\bm{x}}}_H\\
    \label{eq:HDV_CBF}
    &+ \dfrac{\partial b_{i,H}(\bm{x}_i,\Bar{\bm{x}}_H+\bm{e})}{\partial\bm{e}}\dot{\bm{e}} + \alpha(b_{i,H}(\bm{x}_i,\Bar{\bm{x}}_H+\bm{e})) \geq 0.
\end{align}}

\noindent where $\dot{\bm{x}}_i,\dot{\Bar{\bm{x}}}_H$ are described by dynamics \eqref{eq:CAVs_dynamics} and \eqref{eq:hdv_nominal}, respectively, and $b_{i,H}(\bm{x}_i,\Bar{\bm{x}}_H+\bm{e})$ is given by \eqref{eq:safety_constraints_original}.
The only unknown terms left in \eqref{eq:HDV_CBF} are $\bm{e}$ and $\dot{\bm{e}}$, which can be evaluated online directly, i.e., using  (\ref{eq:error}), with $\dot{\bm{e}}$ given by direct measurements of the actual state derivative. Therefore, the satisfaction of \eqref{eq:HDV_CBF} implies the satisfaction of safety constraint $b_{i,H}(\bm{x}_i,\bm{x}_H)\geq 0$ even if the dynamics of $\bm{x}_H$ is unknown to CAVs, as shown in \cite{xiao2022event}. 

However, it is still challenging to solve OCP (\ref{eq:ocp}) since the terminal time $t_f$ is unknown to the controller: in mixed traffic, the uncontrollable and unpredictable behavior of HDVs makes it difficult to properly pre-determine the terminal time $t_f$. 
To overcome this issue, we transform (\ref{eq:ocp}) into a sequence of quadratic programs (QPs), and terminate the lane-changing process by checking $C$'s lateral position with the target lane center at each time step $t_k,k=1,2,...$. Another important step is to update $f_a(\Bar{\bm{x}}_H)$ of the adaptive dynamics \eqref{eq:hdv_nominal} at each $t_k$ with
\begin{equation}
\label{eq:update_hdv}
    f_a(\Bar{\bm{x}}(t^+_k))=f_a(\Bar{\bm{x}}(t^-_k))+\sum_{i=0}^k\dot{\bm{e}}(t_i),
\end{equation}
where $t^+_k,t^-_k$ denote the instants right after and before $t_k$. In this way, we always have the measurements such that $\bm{e}(t_k)=\bm{0}$ and $\dot{\bm{e}}(t_k^+)$ close to 0 at $t_k$ by setting $\Bar{\bm{x}}_H(t_k) = \bm{x}_H(t_k)$. This reduces the number of events (introduced later) to solve the QP and reduce the conservativeness of CAVs.


\subsection{Transform OCP to CBF-based QPs}
In a lane-changing maneuver, we define adaptively updated dynamics for the HDV traveling in the fast lane in the form:
\begin{equation}
\label{eq:nominal_hdv_dynamics}
    \underbrace{\left[\begin{array}{c}
    \dot{\Bar{x}}_H \\
    \dot{\Bar{y}}_H \\
    \dot{\Bar{\theta}}_H \\
    \dot{\Bar{v}}_H   \end{array}\right]}_{\bm{\dot{\Bar{x}}}_H(t)}
    =
    \underbrace{\left[\begin{array}{c}
    \Bar{v}_H \cos \Bar{\theta}_H + h_x(t) \\
    \Bar{v}_H \sin \Bar{\theta}_H + h_y(t) \\
    \Bar{v}_H/L_w + h_{\theta}(t) \\
    h_v(t)    \end{array}\right]}_{f_a\left(\bm{\Bar{x}}_H(t)\right)}
\end{equation}
where $\Bar{\bm{x}}_H = [\Bar{x}_H(t),\Bar{y}_H(t),\Bar{\theta}_H(t),\Bar{v}_H(t)]^T$ represent the estimated longitudinal position, lateral position, heading angle, and speed of the HDV, respectively. Note that $h_x(t),h_y(t),h_{\theta}(t),h_v(t)\in\mathbb{R}^n$ denote the adaptive terms to approximate the real HDV dynamics, where we set $h_j(t_0)=0$ for all $j\in\{x,y,\theta,v\}$.

\begin{rem}
    Different from the general form defined in \eqref{eq:hdv_nominal}, the control input in \eqref{eq:nominal_hdv_dynamics} is implicitly contained in the adaptive terms $h_j(t),j\in\{x,y,\theta,v\}$ when updating the dynamics. The reason is that we cannot directly control the HDV, and observe that CAVs only need HDV states to keep a safe distance. Estimating $h_j(t)$ is equivalent to indirectly estimating $\bm{u}_H$. 
\end{rem}

\textbf{Derive CBF Constraints:} Firstly, we derive the CBF constraints to make sure the speed and control constraint \eqref{eq:uv_constraints}, position constraint \eqref{eq:ylimits}, and safety constraints \eqref{eq:safety_constraints_original} in OCP \eqref{eq:ocp} are satisfied at all times. In order to illustrate how to transform an inequality constraint into a CBF constraint, we consider the safety constraint $b_{C,H}$ between vehicles $C$ and $H$ in \eqref{eq:bCH} as an example in which the unknown HDV dynamics are also included. Computing the Lie derivatives of $b_{C,H}(\bm{x}_C,\bm{x}_H)$ and considering the existence of a state error similar to \eqref{eq:HDV_CBF}, we obtain a new constraint which is linear in the control input and takes the form:
\begin{equation}
\small
L_fb_{C,H}(\bm{x}_C,\!\bm{x}_H)+L_gb_{C,H}(\bm{x}_C,\!\bm{x}_H)\bm{u}_C
\label{eq:CH_CBF}
+\alpha_1(b_{C,H}(\bm{x}_C,\!\bm{x}_H) )\geq 0
\end{equation}
where 
\begin{subequations}
\small
\begin{align}
\nonumber
    &L_fb_{C,H}(\bm{x}_C,\bm{x}_H)=L_fb_{C,H}(\bm{x}_C,\Bar{\bm{x}}_H+\bm{e})\\
    \nonumber
    &= \dfrac{2(x_C-\Bar{x}_H-e_x)}{a_C^2}(v_Ccos\theta_C-\Bar{v}_Hcos\Bar{\theta}_H-h_x-\dot{e}_x)\\
    &+\dfrac{2(y_C-\Bar{y}_H-e_y)}{b_C^2}(v_Csin\theta_C-\Bar{v}_Hsin\Bar{\theta}_H-h_y-\dot{e}_y),\\
    \nonumber
    &L_gb_{C,H}(\bm{x}_C,\bm{x}_H)=L_gb_{C,H}(\bm{x}_C,\Bar{\bm{x}}_H+\bm{e})\\
    \nonumber
    & =[L_{g_{u}}b_{C,H}(\bm{x}_C,\bm{x}_H),L_{g_{\phi}}b_{C,H}(\bm{x}_C,\bm{x}_H)]\\
    \label{eq:L_gCH}
    & = 2v_C[-1,\dfrac{-(x_C\!-\!\Bar{x}_H\!-\!e_x)}{a_C^2}sin\theta_C+\dfrac{(y_C\!-\!\Bar{y}_H\!-\!e_y)}{b_C^2}cos\theta_C],\\
    & \bm{u}_C = [u_C,\phi_C]^T,\\
    \label{eq:CBF_classK}
    & \alpha_1(b_{C,H}(\bm{x}_C,\bm{x}_H)) = k_1b_{C,H}(\bm{x}_C,\Bar{\bm{x}}_H+\bm{e}).
\end{align}
\end{subequations}
where $\bm{e}=[e_x,e_y,e_{\theta},e_v]^T$ denotes the state error.
The class $\mathcal{K}$ function is chosen to be linear in \eqref{eq:CBF_classK} with constant $k_1$. Each of the constraints in \eqref{eq:uv_constraints}, \eqref{eq:ylimits}, and \eqref{eq:safety_constraints_original} can be written in the form $b_n(\bm x)$, $n\in\{1,2,...,N\}, \bm x=\{\bm x_1, \bm x_C, \bm x_H, \bm x_U\}$, where $N$ is the number of constraints, and we can always apply the CBF method to map a constraint $b_n(\bm x)$ to a new CBF constraint for vehicles $i\in\{1,C,H,U\}$ by using the general expression \eqref{eq:CBF_def}
\begin{align}
\label{eq;general_CBFs}
L_fb_n(\bm x)+L_gb_n(\bm x)\bm{u}+\alpha_n(b_n(\bm x)) \geq 0,
\end{align}
with $\bm u = \{\bm u_1, \bm u_C\}$. We omit further details here. 

\textbf{Derive CLF Constraints:} In addition to CBFs used for hard constraints, we use Control Lyapunov Functions (CLFs) associated with the terminal costs in \eqref{eq:ocp} to achieve lane-keeping and to minimize speed deviations from the desired speed $v_d$ for CAVs. Setting $V_1(\bm{x}_C(t))=(v_C(t)-v_d)^2$, $V_2(\bm{x}_1(t))=(v_1(t)-v_d)^2$, $V_3(\bm{x}_C(t))=(y_C(t)-l)^2$, and $V_4(\bm{x}_1(t))=(y_1(t)-l)^2$, the CLF constraints can be expressed as

\begin{subequations}
\small
\label{eq:CLFs}
    \begin{align} 
        & 2(v_C(t)-v_d)u_C(t) + m_1(v_C(t)-v_d)^2 \leq \delta_1(t),\\
        & 2(v_1(t)-v_d)u_1(t)+m_2(v_1(t)-v_d)^2 \leq \delta_2(t),\\
        & 2(y_C(t)\!-\!l)(v_Csin\theta_C\!+\!v_Ccos\theta_C\cdot\phi_C)\!+\!m_3(y_C(t)\!-\!l)^2\!\! \leq \!\delta_3(t),\\
        & 2(y_1(t)-l)(v_1sin\theta_1+v_1cos\theta_1\cdot\phi_1)+m_4(y_1(t)-l)^2 \leq \delta_4(t).
    \end{align}
\end{subequations}
where $m_1,m_2,m_3,m_4$ correspond to the $c_3$ in the CLF constraint (\ref{eqn:clf}), and $\delta_1(t),\delta_2(t),\delta_3(t),\delta_4(t)$ are controllable variables to treat \eqref{eq:CLFs} as soft constraints.

\textbf{Time Discretization:} Finally, we discretize time and let $t_k$, $k=1,2,...$ be the time instants when $C$ solves the QPs. The event time $t_k$ is to be determined (shown in the next subsection). The OCP \eqref{eq:ocp} can be transformed into a sequence of QPs (each solved at the $k$th time step) as follows:
\begin{align}
\label{eq:qp}
    \min_{u_i(t_k),\delta_j(t_k)} \sum_{i=1,C}\alpha_{u_i} u_i^2(t_k) + \sum_{j=1}^4 p_j\delta_j(t_k)
\end{align}
subject to  CBF constraints \eqref{eq;general_CBFs} corresponding to constraints (\ref{eq:uv_constraints}), (\ref{eq:ylimits}), (\ref{eq:safety_constraints_original}), and CLF constraints in \eqref{eq:CLFs}. The weights $\alpha_{u_i},i\in\{1,C\}$ and $p_j,j\in\{1,2,3,4\}$ are adjustable, used to provide a relative importance to each corresponding term in the objective function. 

A common way to discretize time is to select a fixed length $\Delta$ for each time interval such that $t_{k+1}=t_k+\Delta$. Then, \eqref{eq:qp} is solved at each time step $t_k$, and the controls $u_i(t_k),\delta_j(t_k)$ are assumed to be constant over each interval $[t_k,t_k+\Delta)$. This scheme is referred to as the \emph{time-driven} approach. However, this approach does not provide a feasibility guarantee for each CBF-based QP, and safety may be violated due to this time discretization, depending on how $\Delta$ is selected, which is a difficult problem in itself.
We implement 
an \emph{event-driven} approach similar to \cite{xiao2022event} for triggering a QP (with associated CBF constraints) so that it is solved when one of several events (as defined next) is detected. We will show that this event-driven scheme provides a safety guarantee for CAVs to perform the lane-changing maneuver even if the real HDV behavior is unknown.

\subsection{Event-driven Control}
Following the time-driven approach to solving \eqref{eq:qp} at each time step $t_k,k=1,2,...$, we cannot guarantee the satisfaction of CBF constraints because the state error $\bm{e}$ and its derivative $\dot{\bm{e}}$ are generally unknown to the solver right after $t_k$. 
As introduced in \cite{xiao2022event}, the key idea of the event-driven approach is to properly define events depending on state errors and their derivatives so that the QPs will be solved at each event-triggered time $t_k$ while the safety CBF constraints remain satisfied during the time interval $[t_k,t_{k+1})$. 

\textbf{Set Bounds for Error:} In order to find a condition to guarantee the satisfaction of CBF constraints for $t\in[t_k,t_{k+1})$, we first assume that the state error and its derivative satisfy
\begin{subequations}
\begin{align}
\label{eq:e_bd}
    &|e_x|\leq w_x,\; ~|e_y|\leq w_y,\; ~|e_{\theta}|\leq w_{\theta},\; ~|e_v|\leq w_v,\\
    \label{eq:edot_bd}
    &|\dot{e}_x|\leq \nu_x,\; ~|\dot{e}_y|\leq \nu_y,\; ~|\dot{e}_{\theta}|\leq \nu_{\theta},\; ~|\dot{e}_v|\leq \nu_v,
\end{align}
\end{subequations}
where $\bm{w}:=[w_x,w_y,w_{\theta},w_v]^T\in\mathbb{R}^4_{\geq 0}$, and $\bm{\nu}:=[\nu_x,\nu_y,\nu_{\theta},\nu_v]^T\in\mathbb{R}^4_{\geq 0}$ are chosen bounds that determine the conservativeness of the framework. Specifically, small bounds introduce less conservativeness with more events, and vice versa. Similarly, we consider the state vector $\bm{x}_i$ for all vehicles $i\in\{1,C,H,U\}$ at time $t_k$, which satisfies the following bounds
\begin{align}
\label{eq:state_bounds}
    \bm{x}_i(t_k)-\bm{s}_i \leq \bm{x}_i(t) \leq \bm{x}_i(t_k)+\bm{s}_i,
\end{align}
where $\bm{s}_i\in\mathbb{R}^4_{\geq 0}$ is a parameter vector similar to error bounds. We denote a set for states of vehicle $i$ that satisfy \eqref{eq:state_bounds} at time $t_k$ as
\begin{align}
\label{eq:set}
    S_i(t_k)=\{\bm{z}_i\in X:\bm{x}_i(t_k)-\bm{s}_i \leq \bm{z}_i \leq \bm{x}_i(t_k)+\bm{s}_i\}
\end{align}
According to Definition \ref{def:Forward_Invariant_Set}, we define a feasible set $C_{i,1}$ for vehicle $i\in\{1,C,H,U\}$ in the following such that all original constraints (\ref{eq:CAVs_dynamics}),~(\ref{eq:uv_constraints}),~(\ref{eq:ylimits}),~(\ref{eq:safety_constraints_original}) are satisfied:
\begin{equation}
    C_{i,1}:=\{\bm{x}_i\in X: b_n(\bm{x})\geq 0,~ n\in{1,2,...,N}\}
\end{equation}
Based on these settings, we can proceed to find a condition that guarantees the satisfaction of all CBF constraints in the time interval $[t_k,t_{k+1})$, which can be done by minimizing each component in \eqref{eq;general_CBFs}. We still take the CBF constraint \eqref{eq:CH_CBF} between vehicles $C$ and $H$ as an example to illustrate the detailed process.

\textbf{Find Robust CBFs:} In \eqref{eq;general_CBFs}, let $L_{f_{\min}}b_{C,H}(t_k)\in\mathbb{R}$ be the minimum value of $L_fb_{C,H}(\bm{x}_C,\bm{x}_H)
+k_1b_{C,H}(\bm{x}_C,\bm{x}_H) $ for the proceeding time interval that satisfies $\bm r \in R(t_k)$ where $\bm r:=[\bm{z}_C, \bm{z}_H, \bm{e}, \dot{\bm{e}}]^T, R(t_k):=\{\bm r: \bm{z}_C\in S_C(t_k),\bm{z}_H\in S_H(t_k),|\bm{e}|\leq \bm{w}, |\dot{\bm{e}}|\leq \bm{\nu},\bm{z}_H+\bm{e}\in C_{H,1}\}$ starting at time $t_k$, i.e.,
\begin{equation}\label{eq:minterm_Lf}
\begin{adjustbox}{width=\linewidth}
\small
    $L_{f_{\min}}b_{C,H}(t_k) = \min\limits_{\bm r \in R(t_k)} 
   L_fb_{C,H}(\bm{z}_C,\bm{z}_H)
+k_1b_{C,H}(\bm{z}_C,\bm{z}_H)$
\end{adjustbox}
\end{equation}
The remaining term $L_gb_{C,H}(\bm{x}_C,\bm{x}_H)\bm{u}_C$ in \eqref{eq:CH_CBF} contains the control input $\bm{u}_C = [u_C,\phi_C]^T$, which complicates the minimization process.
To minimize this term, we need to consider the sign of $\phi_C,u_C$ at time $t_k$. Similar to \eqref{eq:minterm_Lf}, set 
\begin{equation*}
    L_{g_{\min}}b_{C,H}(t_k)=[L_{g_{u\min}}b_{C,H}(t_k),L_{g_{\phi\min}}b_{C,H}(t_k)]
\end{equation*}
as the vector of minimum values of $L_{g}b_{C,H}(\bm{x}_C,\bm{x}_H)$ in \eqref{eq:L_gCH} for the proceeding time interval. Then, we have 
\begin{equation}
\begin{adjustbox}{width=\linewidth}
    $L_{g_{\phi\min}}b_{C,H}(t_k)=
    \begin{cases}
       \min\limits_{\bm r \in R(t_k)} 
   L_{g_{\phi}}b_{C,H}(\bm{z}_C,\bm{z}_H), &\!\! if \;\phi_C(t_k)\geq 0,\\
        \max\limits_{\bm r \in R(t_k)} 
  L_{g_{\phi}}b_{C,H}(\bm{z}_C,\bm{z}_H), &\!\! {otherwise.}
    \end{cases}$
    \end{adjustbox}
\end{equation}
and 
\begin{equation}
\begin{adjustbox}{width=\linewidth}
    $L_{g_{u\min}}b_{C,H}(t_k)=
    \begin{cases}
       \min\limits_{\bm r \in R(t_k)} 
   L_{g_{u}}b_{C,H}(\bm{z}_C,\bm{z}_H), &\!\! if \;u_C(t_k)\geq 0,\\
        \max\limits_{\bm r \in R(t_k)} 
  L_{g_u}b_{C,H}(\bm{z}_C,\bm{z}_H), &\!\! {otherwise.}
    \end{cases}$
    \end{adjustbox}
\end{equation}
where the sign of $\phi_C,u_C$ can be obtained by simply solving the CBF-based QP \eqref{eq:qp} at time $t_k$. Therefore, the condition that guarantees the satisfaction of \eqref{eq:CH_CBF} during $[t_k,t_{k+1})$ is given by
\begin{equation}
\label{eq:event_condition}
    L_{f_{\min}}b_{C,H}(t_k) + L_{g_{\min}}b_{C,H}(t_k)\bm{u}_C \geq 0.
\end{equation}
Similarly, for all CBF constraints with the form \eqref{eq;general_CBFs}, they have to satisfy the condition
\begin{align}
\label{eq:event_CBFs}
    L_{f_{\min}}b_n(t_k)+L_{g_{\min}}b_n(t_k)\bm{u} \geq 0,
\end{align}
where $L_{f_{\min}}b_n(t_k)$ is the minimum value of $L_fb_n(\bm{x})+\alpha_n(b_n(\bm{x}))$, and $L_{g_{\min}}b_n(t_k)\bm{u}$ is the minimum value of $L_gb_n(\bm{x})\bm{u}$ during $[t_k,t_{k+1})$. 

In order to apply the above conditions to the QP \eqref{eq:qp}, we just replace all the CBF constraints \eqref{eq;general_CBFs} by \eqref{eq:event_CBFs}, i.e., 
\begin{align}
\label{eq:qp_event}
    \min_{u_i(t_k),\delta_j(t_k)} \sum_{i=1,C}\alpha_{u_i} u_i^2(t_k) + \sum_{j=1}^4 p_j\delta_j(t_k)
\end{align}
subject to the CAV dynamics \eqref{eq:CAVs_dynamics}, CBF conditions \eqref{eq:event_CBFs}, and CLF constraints in \eqref{eq:CLFs}.

\textbf{Determine Triggering Events:} Based on the above settings, we define three events that specify the conditions for triggering an instance of solving QP \eqref{eq:qp_event}:  \\
\textbf{Event 1:} The measured HDV state error $\bm{e}$ exceeds its bounds $\bm{w}$, i.e., any one inequality in \eqref{eq:e_bd} is about to be violated.\\
\textbf{Event 2:} The measured derivative of the HDV state error $\bm{\dot{e}}$ exceeds its bounds $\bm{\nu}$, i.e., any one inequality in \eqref{eq:edot_bd} is about to be violated.\\
\textbf{Event 3:} The state measurement of vehicle $i,i\in\{1,C,H,U\}$ reaches the boundaries of $S_i(t_k)$.

The first two events can be detected by directly measuring the state error and its derivative, and the third event is detected by monitoring the dynamics of \eqref{eq:CAVs_dynamics}  and \eqref{eq:nominal_hdv_dynamics}. Therefore, the next event-triggered time $t_{k+1},k=0,1,2,...$ is given by 

{\small
\begin{align}
\nonumber
    t_{k+1} = \min \{t>t_k:&|\bm{e}|=\bm{w}\; \text{or}\; |\dot{\bm{e}}| = \bm{\nu}\;  \text{or} \; 
    |\bm{x}_i(t)\!-\!\bm{x}_i(t_k)|=\bm{s}_i\; \\
    \label{eq:tk+1}
     &\text{or}\; |\bm{\Bar{x}}_H(t)-\bm{\Bar{x}}_H(t_k)|=\bm{s}_H\},
\end{align}}

\noindent where $i\in\{1,C,U\}$ in \eqref{eq:tk+1}. Recalling that the real HDV dynamics are unknown, we apply its estimated states $\Bar{\bm{x}}_H$ from \eqref{eq:nominal_hdv_dynamics} to check the next triggered time $t_{k+1}$. The choice of each component of $\bm{s}_i$ in (\ref{eq:state_bounds})
captures the tradeoff between the time complexity and the conservativeness of the approach. A larger magnitude of $\bm{s}_i$ corresponds to a smaller number of trigger times for solving the QPs, which reduces the total computation costs. However, the conditions \eqref{eq:event_condition} and \eqref{eq:event_CBFs} must be satisfied over a longer time interval, which renders the approach more conservative.

\subsection{Lane-Changing Implementation with Unkown HDV Dynamics}
In this section, we describe the process of applying the event-triggered CBFs for CAV $C$ in Fig. \ref{fig:lane_change_process} to complete the lane-changing maneuver safely and optimally. Assuming CAV $C$ starts to change lanes at time $t_0$, we proceed as follows:

\emph{1) Step 1:} Given the initial states $\bm{x}_i(t_0)=[x_i(t_0),y_i(t_0),\theta_i(t_0),v_i(t_0)]^T$, $i\in\{1,C,H,U\}$ for all vehicles, predetermine the error bound parameters $\bm{w}=[w_x,w_y,w_{\theta},w_v]^T$, $\bm{\nu}=[\nu_x,\nu_y,\nu_{\theta},\nu_v]^T$ for the HDV, and set $\bm{s}_i=[s_{i_x},s_{i_y},s_{i_{\theta}},s_{i_v}]^T$ for $S_i$ in \eqref{eq:set}. Define adaptive dynamics \eqref{eq:nominal_hdv_dynamics} for the uncontrollable HDV involved in the maneuver.

\emph{2) Step 2:} Measure $\bm{x}_i$ for all vehicles and $\dot{\bm{x}}_H$ for the HDV at $t_k$ (where $k$ starts from 0), and get the state error $\bm{e}$ and its derivative $\dot{\bm{e}}$ for the HDV based on \eqref{eq:error}.
Then, evaluate the CBF condition in \eqref{eq:event_CBFs} for all constraints in (\ref{eq:uv_constraints}),~(\ref{eq:ylimits}),~(\ref{eq:safety_constraints_original}) similar to the calculations in \eqref{eq:minterm_Lf}-\eqref{eq:event_condition}. Form CLF constraints in \eqref{eq:CLFs}, and transform the original OCP \eqref{eq:ocp} for the CAVs to a series of CBF-based QPs in \eqref{eq:qp_event}.

\emph{3) Step 3:} 
Solve the CBF-based QP \eqref{eq:qp_event} at $t_k$ to obtain the optimal control $u_1^*(t_k),u_C^*(t_k)$ for CAVs 1 and $C$, respectively, and apply them after $t_k$. Meanwhile, monitor the lateral position of CAV $C$ to check if $C$ completes its lane-changing maneuver. In particular, for $t\geq t_k$, if $|y_C(t)-l|\leq \epsilon$, where $\epsilon>0$ is a tolerable error, then $C$ achieves its target lane and we terminate the process.

\emph{4) Step 4:}
If $|y_C(t)-l|> \epsilon$, determine the next event-triggered time instant $t_{k+1},k=0,1,2,...$ based on \eqref{eq:tk+1}. At the next time $t_{k+1}$, synchronize the estimated HDV dynamics \eqref{eq:nominal_hdv_dynamics} by using \eqref{eq:update_hdv}:
\begin{equation}
\label{eq:h_i_update}
h_j(t^+)=h_j(t^-)+\sum\limits_{i=0}^{k}\dot{e}_j(t_i),~~j\in\{x,y,\theta,v\}     
\end{equation}
Let $k=k+1$, and return to \emph{Step 2}.

\emph{5) Step 5:} If a maximum prespecified allowable maneuver time $T_f$ is reached and we still have $|y_C(T_f)-l|> \epsilon$, then we abort the maneuver for $C$.

We summarize the entire process in Algorithm. \ref{alg:event}.

\begin{algorithm}[hpt]
    \caption{Implementation of event-triggered CBFs in Lane-changing maneuvers}
    \SetKwInOut{Input}{input}
    \SetKwInOut{Output}{output}
    \Input{Initial States $\bm{x}_i(t_0),i=1,C,H,U$, measurements of $\bm{\dot{x}}_H(t_0)$, HDV estimated dynamics \eqref{eq:nominal_hdv_dynamics}, error bounds of $\bm{w},\bm{\nu},\bm{s}_i$, settings for QP \eqref{eq:qp_event}.}
    \Output{Event time $t_k,k=1,2,...$, the optimal control $u_1^*(t_k),u_C^*(t_k)$ for CAVs, terminal time $t_f$.}
    \SetKwBlock{Beginn}{beginn}{ende}
    \Begin{
        $k=0,t_k=t_0.$\\
        \While{$t_k\leq T_f$}{
            Measure $\bm{x}_H$ and $\bm{\dot{x}}_H$ at time $t_k$,\\
            Synchronize the HDV dynamics \eqref{eq:nominal_hdv_dynamics} by \eqref{eq:h_i_update},\\
            Evaluate the CBF condition \eqref{eq:event_CBFs} for all constraints (\ref{eq:uv_constraints}),~(\ref{eq:ylimits}),~(\ref{eq:safety_constraints_original}) similar to \eqref{eq:minterm_Lf}-\eqref{eq:event_condition},\\
            Solve QP \eqref{eq:qp_event} at $t_k$ and get $u_1^*(t_k),u_C^*(t_k)$.\\
        \While{$t<T_f$}{
            Apply $u_1^*(t_k),u_C^*(t_k)$ for $t>t_k$.\\
            Measure $\bm{x}_{1,C,U,H}$ and $\bm{\dot{x}}_H$,\\
            \If{$|y_C(t)-l|\leq \epsilon$}{
            maneuver completed, terminate the process, $t_f=t$.
            } 
            Evaluate $t_{k+1}$ by \eqref{eq:tk+1}.\\
            \If{$t_{k+1}$ is found}{
                   $k \leftarrow k+1$, break.\\
                } 
        }
    }
    \If{$|y_C(T_f)-l|>\epsilon$}{
                   Abort maneuver.\\
                } 
    }
    \label{alg:event}
\end{algorithm}

\section{SIMULATION RESULTS}
\begin{figure*}[hpbt]
    \centering   
    \begin{subfigure}{0.32\linewidth}
    \centering 
      \includegraphics[width=\linewidth]{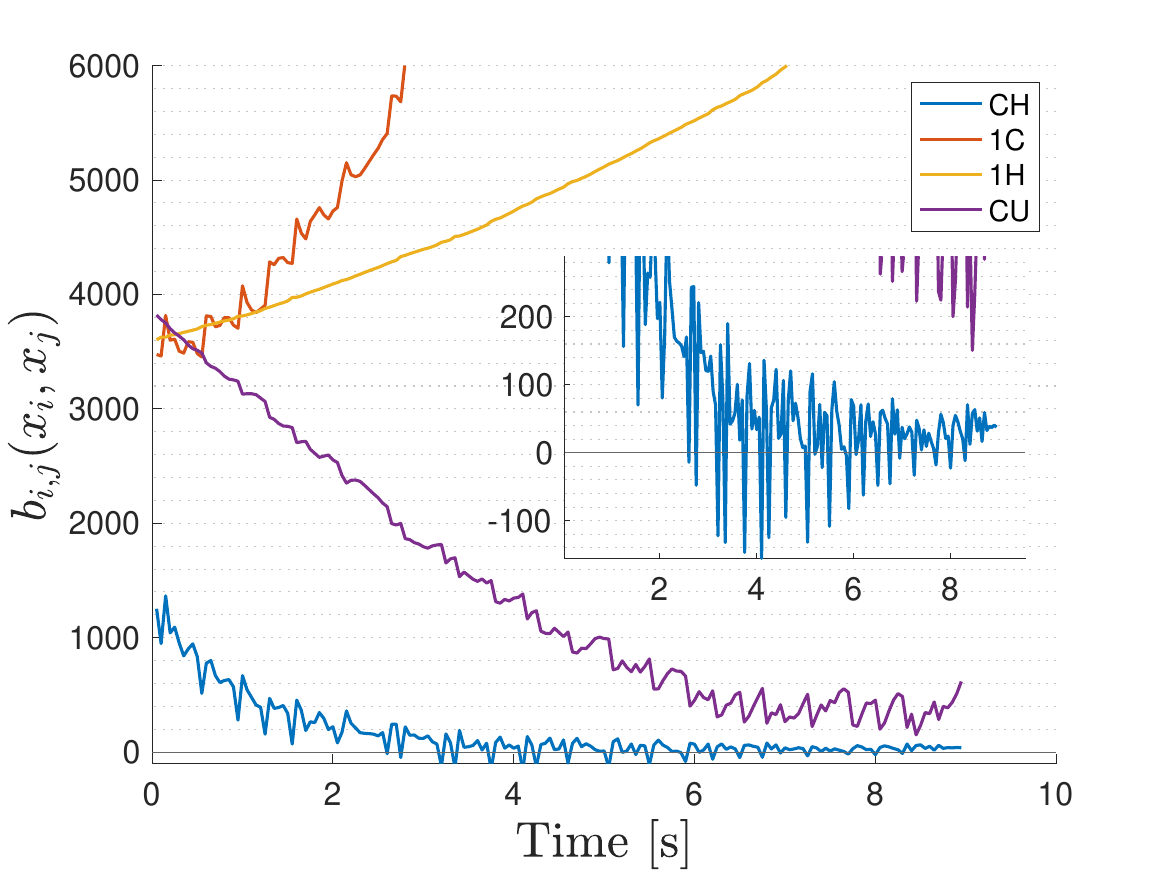}  
      \caption{\centering{Case 1: Time-driven approach with known HDV dynamics}}
      \label{fig:case1}
    \end{subfigure}   
    \begin{subfigure}{0.32\linewidth}
    \centering 
      \includegraphics[width=\linewidth]{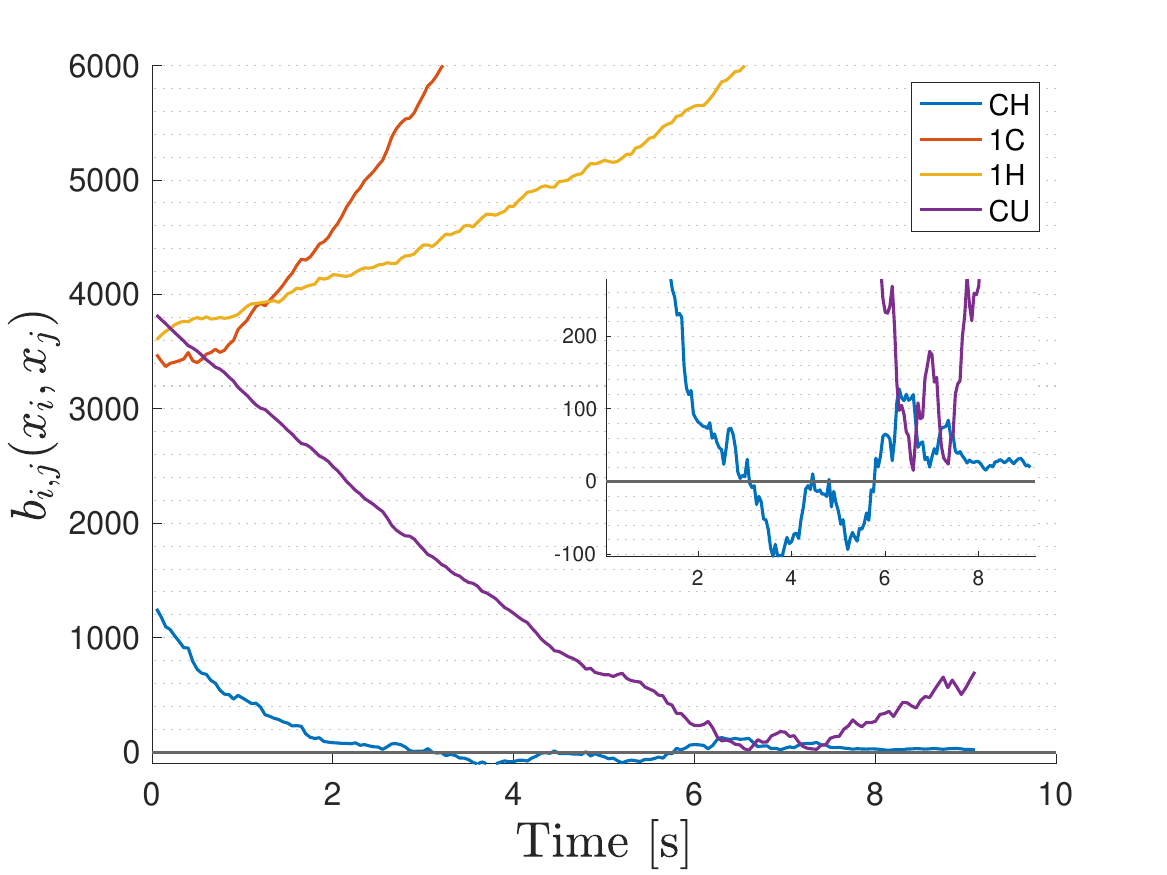}  
      \caption{\centering{Case 2: Time-driven approach with unknown HDV dynamics  }}
      \label{fig:case2}
    \end{subfigure}
    \begin{subfigure}{0.32\linewidth}
    \centering 
      \includegraphics[width=\linewidth]{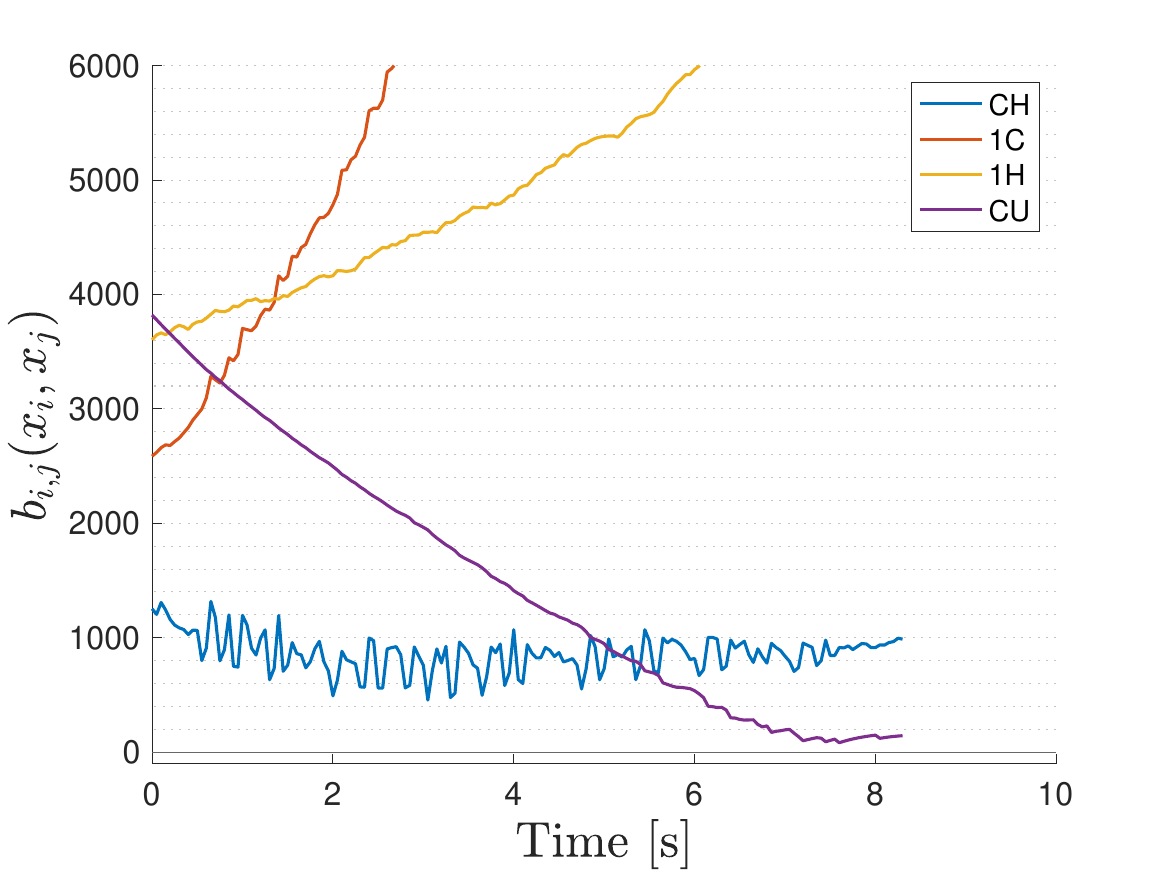}  
      \caption{\centering{Case 3: Event-driven approach with unknown HDV dynamics}}
      \label{fig:case3}
    \end{subfigure}  
    \caption{\small Safety with time-driven and event-driven CBFs. $b_{i,j}(\bm{x}_i,\bm{x}_j)$ denotes the value of the CBF between vehicles $i$ and $j$, where $(i,j)\in\{(C,H),(1,C),(1,H),(C,U)\}$. $b_{i,j}(\bm{x}_i,\bm{x}_j)\geq 0$ denotes safety guarantees (Case 3, not Cases 1 and 2).}
    \label{fig:time_event_comparison}  
\end{figure*}
This section provides simulation results illustrating the optimal lane-changing trajectories for each CAV with safety guarantees in mixed traffic, even though the HDV dynamics are unknown to CAVs. We emphasize that the controller can be implemented to monitor the HDV's behavior and respond to HDVs in real-time. We test our framework by allowing human drivers to manually operate virtual vehicles through a MATLAB interface, and the results show that CAV $C$ can always update its control to avoid collisions and successfully perform a safe maneuver. Our simulation setting is that of Fig. \ref{fig:lane_change_process}. Vehicle $U$ is assumed to travel with constant speed $v_U = 20 m/s$ all the time 
(this is not needed in the overall approach).
The allowable speed range for CAVs is $v\in[15,35] m/s$, and the acceleration of vehicles is limited to $\bm u\in[(-7,-\pi/4),(3.3,\pi/4)] m/s^2$. The desired speed $v_d$ for the CAVs is considered as the traffic flow speed, which is set to $30 m/s$. To guarantee safety, in designing the size of the ellipse in \eqref{eq:safety_distance} as a safe region, we set the parameters $a_C=a_1=0.6$ as the reaction time of CAVs, and $b_1=b_C=0.1$ to let the minor axis approximate the lane width $l=4 m$. The maximum allowable maneuver time is set at $T_f=15 s$. 
The real HDV dynamics are \textbf{unknown} to the controller and expressed as:
\begin{equation}
\label{eq:hdv_real}
\begin{adjustbox}{width=\linewidth}
    $\left[\begin{array}{c}
    \dot{x}_H \\
    \dot{y}_H \\
    \dot{\theta}_H \\
    \dot{v}_H   \end{array}\right]$
    =
    $\left[\begin{array}{c}
    v_H \cos \theta_H \cdot\sigma_1 \\
    v_H \sin \theta_H\cdot \sigma_2 \\
    0 \\
    0    \end{array}\right]+$
    $\left[\begin{array}{cc}
    0 & -v_H \sin \theta_H \\
    0 & v_H \cos \theta_H \\
    0 & v_H / L_w \\
    1 & 0    \end{array}\right]
    \left[\begin{array}{l}
    u_H \\
    \phi_H
    \end{array}\right]
    +$
    $\left[\begin{array}{c}
    \varepsilon_1\\
  \varepsilon_2\\
  \varepsilon_3\\
  \varepsilon_4  \end{array}\right]$
  \end{adjustbox}
\end{equation}
where $\bm u_H$ is either a random policy or controlled by a human player. $\sigma_1,\sigma_2$ denote two random processes with uniform pdfs over the interval $[0.9,1.1]$,
and $\varepsilon_1\in[-0.7,0.7],\varepsilon_2\in[-0.5,0.5],\varepsilon_3\in[-0.5,0.5],\varepsilon_4\in [-0.7,0.7]$ are disturbances. The initial states of vehicles at time $t_0=0$ are given as $\bm{x}_C(t_0)=[20m,0m,0rad,25m/s]^T$, $\bm{x}_1(t_0)\!\!\!=\!\!\![50m,4m,0rad,29m/s]^T$, $\bm{x}_H(t_0)\!\!=\!\![10m,4m,0m,28m/s]^T$\!,\! $\bm{x}_U(t_0)\!\!=\!\![60m,0m,0rad,20m/s]^T$. The weights in QP \eqref{eq:qp_event} are set as $\alpha_{u_1}=\alpha_{u_C}=1$, $p_1=p_2=p_4=1,p_3=100$. The initial states of the adaptive HDV dynamics \eqref{eq:nominal_hdv_dynamics} are set as $\bm{\bar{x}}_H(t_0)=\bm{x}_H(t_0)$, and the adaptive terms at $t_0$ satisfy $h_x(t_0)=h_y(t_0)=h_{\theta}(t_0)=h_v(t_0)=0.$ 

The parameters $\bm{s}_i$ in the feasible set $S_i$ are $\bm{s}_i=[0.01m,0.005m,0.01rad,1m/s]$ for all vehicles $i\in\{1,C,H,U\}$. The bound for state error $\bm{e}$ and its derivative $\bm{\dot{e}}$ are given as $\bm{w}=[0.2m,0.1m,0.1rad,1m/s]$, $\bm{\nu}=[0.5m/s,0.2m/s,0.1rad/s,1m/s^2]$. The allowable error to terminate the maneuver is $\epsilon=0.3m$. The numerical solutions to the QPs are obtained using an interior point optimizer (IPOPT) on an Intel(R) Core(TM) i7-8700 3.20GHz. The computation times for time-driven and event-driven approaches are $1.5ms$ and $24.0ms$, respectively.

\subsection{Comparison between Time and Event Driven Approach}
\label{sec:event_time_comparison}
Based on the above settings, we compare our event-triggered approach in solving CBF-based QPs \eqref{eq:qp_event} with unknown HDV dynamics to a time-driven approach. Set the discretized time interval $\Delta=0.05s$. Due to inter-sampling effects on system performance when applying a time-driven approach, we consider three cases to test the effectiveness of the event-driven approach in implementing the lane-changing problem. The HDV policy is set to be random, satisfying $u_H(t_k)\in[-1.7,1.7]m/s^2,\phi_H(t_k)\in[-0.2\pi,0.2\pi]rad,k=0,1,2,...$. 

\emph{Case 1: Apply time-driven approach with known HDV dynamics}. Assume HDV dynamics \eqref{eq:hdv_real} with disturbances are known to CAVs when we apply a time-driven approach.

\emph{Case 2: Apply time-driven approach with unknown HDV dynamics.} Assume the HDV dynamics are unknown to CAVs. HDV states have to be estimated at each time step $t_k=t_0+k\Delta$, $k=0,1,2,...$.

\emph{Case 3: Apply event-driven approach with unknown HDV dynamics.} Assume HDV dynamics are unknown to CAVs. HDV states have to be estimated at each time step $t_k,k=0,1,2,...$ given by \eqref{eq:tk+1}.

The simulation results are shown in Fig. \ref{fig:time_event_comparison}, where the x-axis denotes the simulation time and the y-axis denotes the value of safety constraint $b_{i,j}$ in \eqref{eq:safety_distance}. $b_{i,j}<0$ represents a violation of the safety constraint between vehicles $i$ and $j$. In Fig. \ref{fig:time_event_comparison}, the distances between vehicles 1 and $C$ (red curve), vehicles $1$ and $H$ (yellow curve) keep increasing. 
The two constraints about to be violated are $b_{C,U}$ (purple curve) and $b_{C,H}$ (blue curve). From Fig. \ref{fig:case1}, even if the HDV dynamics are assumed to be known to CAV $C$, we still have $b_{C,H}<0$ at some points, which means the distance between vehicles $C$ and $H$ is less than the safe distance.
Similar results occur in Fig. \ref{fig:case2}, where $b_{C,H}$ (blue curve) is below 0 at some points, violating safety during the maneuver. Safety is not guaranteed even with state synchronization under the time-driven approach. In Fig. \ref{fig:case3}, all curves are above 0, implying safety guarantees for all vehicles during the lane-changing maneuver.

\subsection{Human Case Studies}
\begin{figure}[hpbt]
    \centering   
    \begin{subfigure}{0.23\linewidth}
    \centering 
      \includegraphics[width=\linewidth]{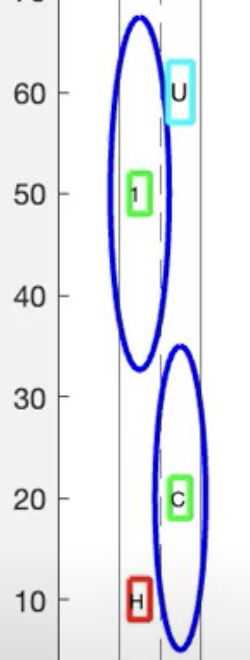}  
      \caption{\centering{Snapshot $t = 0s$}}
    \end{subfigure}   
    \begin{subfigure}{0.23\linewidth}
    \centering 
      \includegraphics[width=\linewidth]{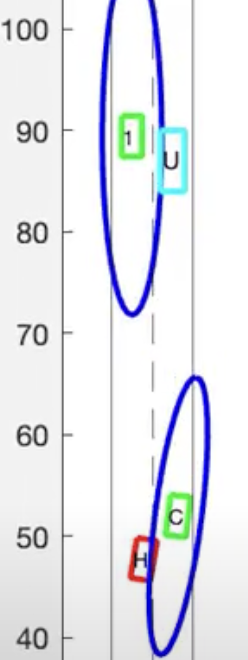}  
      \caption{\centering{Snapshot $t = 1s$ }}
    \end{subfigure}
    \begin{subfigure}{0.23\linewidth}
    \centering 
      \includegraphics[width=\linewidth]{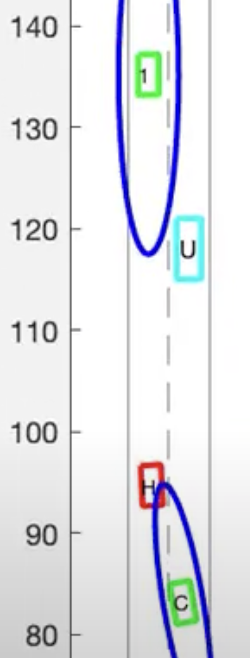}  
      \caption{\centering{Snapshot $t = 3s$}}
    \end{subfigure} 
    \begin{subfigure}{0.23\linewidth}
    \centering 
      \includegraphics[width=\linewidth]{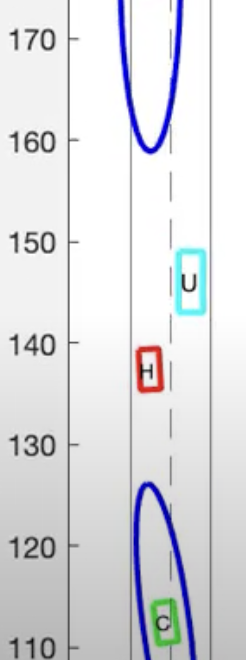}  
      \caption{\centering{Snapshot $t = 5s$}}
    \end{subfigure}  
    \caption{\small Snapshots of human study with an aggressive player using the proposed framework. The red HDV is controlled by an aggressive human player, the green vehicles are CAVs and cyan vehicle is a blocking vehicle. Blue ellipses denote safe regions. Safety is guaranteed between the HDV and CAVs with human-in-the-loop. }
    \label{fig:human}  
\end{figure}

The simulation results in Sec. \ref{sec:event_time_comparison} illustrate the effectiveness of the event-driven approach to guarantee safety in lane-changing maneuvers with random HDV policies. In this section, we further introduce human control in the framework, from which the human driver's aggressiveness will affect CAVs' responses. We have drivers perform \emph{aggressive, hesitant, and conservative} driving behaviors to test the proposed approach through the merging point, safety satisfaction, maneuver time, and energy consumption. The aggressive driver exhibits acceleratory tendencies irrespective of surrounding vehicles, refusing to yield right-of-way. In contrast, the conservative driver prioritizes safety contingencies above all else, readily yielding to proximate vehicles. Finally, the hesitant driver is characterized by driving uncertainty, responding to surrounding vehicles with sudden, inconsistent accelerations and decelerations. Taking the aggressive performance as an example, snapshots of how the maneuver evolves are shown in Fig. \ref{fig:human}. The simulation results for three types of driving players are summarized in Table. \ref{tab:human_study}. Given the safety constraint that is about to be violated is $b_{C,H}$ between vehicles $C$ and $H$ (from the results in Fig. \ref{fig:time_event_comparison}), the column ``Safety'' in Table. \ref{tab:human_study} is defined as the minimum value of $b_{C,H}(t_k),k=0,1,2,...$ during the entire maneuver. 

\begin{table}[]
\resizebox{\linewidth }{!}{%
\begin{tabular}{cccccc}
\toprule
\multirow{2}{*}{\begin{tabular}[c]{@{}c@{}}Human\\ Driver  Type\end{tabular}} & \multicolumn{2}{c}{Times} & \multirow{2}{*}{Safety} & \multirow{2}{*}{\begin{tabular}[c]{@{}c@{}}Terminal \\Time  $t_f [s]$\end{tabular}} & \multirow{2}{*}{\begin{tabular}[c]{@{}c@{}}Energy \end{tabular}} \\ 
\cline{2-3}
& A-HDV       & B-HDV      &                         &                                                                                     &                                                                                \\ \midrule
Aggressive                                                                           & 0                  & 10              & 627.7                & 3.4$\pm$0.3                                                                              & 27.1$\pm$25.9                                                                         \\
Hesitant                                                                               & 5                  & 5               & 521.2                 & {\color{red}8.8$\pm$1.4}                                                                               & {\color{red}63.2$\pm$46.2}                                                                       \\
Conservative                                                                         & 10                 & 0               & 575.6                & 3.8$\pm$0.7                                                                               & 18.8$\pm$17.0                                                                         \\ \bottomrule
\end{tabular}}
\caption{\small Performance of CAV $C$ under different human driver types. ``A-HDV" and ``B-HDV" represent merging ahead of HDV and behind HDV, respectively. ``Safety" denotes the minimum value of $b_{C,H}$ during the entire maneuver in the repeated 10 times.}
\label{tab:human_study}
\end{table}

Table \ref{tab:human_study} shows that if the human driver is aggressive, $C$ is always conservative and chooses to merge behind the HDV. On the contrary, if the human driver is conservative, then it is safe for $C$ to behave aggressively and merge ahead of the HDV. If the human driver is hesitant, the merging point varies and depends on the real-time traffic conditions. Note that all values in the Safety column are positive, which indicates no safety constraint is ever violated under the proposed event-driven approach. Moreover, considering the maneuver time $t_f$ in view of energy consumption, we notice that when the driver's intention is explicit, i.e., the human driver is aggressive or conservative, $C$ can respond and merge quickly by adapting to the HDV's behavior: the average maneuver time is $3.4s$ and $3.8s$, respectively, with corresponding energy consumptions 27.1 and 18.8. However, if the human driver performs hesitantly, the driver intention is not clear to CAV $C$, so that it always travels in a conservative manner with a longer average maneuver time of $8.8s$, and higher energy consumption of $63.2$. This motivates exploring an optimal way to evaluate human characteristics in advance so that $C$ can make decisions earlier, hence improving its performance. The videos for three types of drivers (human players) can be found in {\small \color{blue}\url{https://drive.google.com/drive/folders/1JQQ0mRMX35bEV6wsbPHmoM6UdOLMW2Dv?usp=sharing}}.

\section{CONCLUSIONS}

This paper proposes a robust framework for safe human interactions in mixed traffic, in which case the connected and automated vehicles can always guarantee safety with respect to human driven vehicles. This framework is mainly based on the real-time estimation of the human driven vehicle dynamics and control policy and the incorporation of such estimations into the event-triggered control barrier functions. Simulation results in mixed traffic highway merging with different types of human drivers have demonstrated the effectiveness and robustness of the proposed framework in guaranteeing the safety of all the vehicles. Future work will focus on estimating the characteristics of human drivers and seeking a more efficient performance for CAV maneuvers.



\bibliographystyle{IEEEtran}

\bibliography{cmp}

\end{document}